\documentclass{blois}

\def\Journal#1#2#3#4{{#1} {\bf #2}, #3 (#4)}

\def\PRL{\em Phys. Rev. Lett.}

\def\be{\begin{equation}}
\def\ee{\end{equation}}
\def\bea{\begin{eqnarray}}
\def\eea{\end{eqnarray}}

\usepackage{xspace}
\usepackage{subcaption}
\usepackage{enumitem}
\setlist[itemize]{noitemsep, topsep=0pt}

\def\qz{\mbox{CUORE-0}\xspace}

\def\holder{\emph{Holder}\xspace}

\def\crystal{\emph{Crystals}\xspace}

\def\mspecone{$\mathcal{M}_1$\xspace}
\def\mspectwo{$\mathcal{M}_2$\xspace}

\def\summspectwo{$\Sigma_2$\xspace}

\def\tect{$^{130}$Te\xspace}
\def\teod{TeO$_2$\xspace}
\def\bb{$\beta\beta$\xspace}
\def\bbd{$2\nu\beta\beta$\xspace}
\def\bbz{$0\nu\beta\beta$\xspace}

\def\Td{$T_{1/2}^{2\nu}$\xspace}

\def\u{$^{238}$U\xspace}
\def\th{$^{232}$Th\xspace}
\def\kq{$^{40}$K\xspace}
\def\co{$^{60}$Co\xspace}

\def\mspecone{$\mathcal{M}_1$\xspace}
\def\mspectwo{$\mathcal{M}_2$\xspace}
\def\summspectwo{$\Sigma_2$\xspace}

\newcommand{\Photo}{}

\begin{document}
\vspace*{4cm}
\title{CUORE-0 background analysis and evaluation of $^{130}$Te $2\nu\beta\beta$ decay half-life}

\author{Davide Chiesa on behalf of the CUORE collaboration }
\address{University and INFN of Milano-Bicocca, Milano I-20126 - Italy}

\maketitle\abstracts{CUORE is a bolometric experiment that will search for the Neutrinoless Double Beta decay of \tect. 
\qz is a single CUORE-like tower that was run between 2013 and 2015 to test the performance of the CUORE experiment. 
In this proceeding we present the results of the model developed to disentangle and quantify the background sources that combine to form the \qz energy spectrum. 
We use detailed Geant4-based simulations and a Bayesian fitting algorithm to reconstruct the experimental data and evaluate the activities of the background sources.
A direct outcome of this analysis is the measurement of the \tect \bbd decay half-life, of which we provide a preliminary evaluation.
}

\small

\section{Introduction}

The detection of neutrino oscillations proved that neutrinos are massive particles. However, some questions are still open about their absolute mass scale and their nature, i.e. if neutrinos are Dirac or Majorana particles. The observation of the Neutrinoless Double Beta (\bbz) decay~\cite{Furry} would allow to answer both questions. The Double Beta (\bb) decay is a rare nuclear process in which an (A,~Z) nucleus decays into its (A,~Z+2) isobar with the simultaneous emission of two electrons. In the Standard Model, \bb decay is an allowed transition if two (anti-)neutrinos are emitted to conserve lepton number. This decay is called Two-Neutrino Double Beta (\bbd) decay and is the slowest process ever directly observed~\cite{ExoKz}. The same transition through a channel in which no neutrinos are emitted (the \bbz decay) 
violates lepton number conservation and is possible only if neutrinos are massive Majorana particles. 
The \bb decay is detected measuring the kinetic energy of the two emitted electrons. The signature of the \bbz decay is a peak at the Q-value of the transition, while the \bbd decay produces a continuum spectrum in the detector, because neutrinos take away some part of the energy.

The Cryogenic Underground Observatory for Rare Events (CUORE) experiment~\cite{Cuore} is the latest and most massive bolometric detector designed to search for the \bbz decay of \tect (34\% isotopic abundance, Q-value at 2528~keV). 
The CUORE detector is composed by 988 \teod bolometers, arranged in a structure of 19 towers, for a total mass of 741~kg (206~kg of \tect). CUORE is now in its final commissioning phase at Laboratori Nazionali del Gran Sasso (LNGS, Italy) and 
will start data taking by the end of 2016.

\qz is the first tower from the CUORE detector assembly line and it was operated at LNGS between 2013 and 2015~\cite{Q0-detector}. In addition to being a competitive \bbz experiment~\cite{Q0}, \qz was a proof of concept of CUORE in all stages from the assembly line to the DAQ and analysis framework. Last but not least, the reconstruction of the background sources responsible for \qz counting rate enabled us to verify that the necessary background requirements for CUORE are fulfilled. 


\section{\qz detector, experimental data and Monte Carlo simulations}

\begin{figure}[b!]
\begin{center}
\captionsetup[subfigure]{labelformat=empty}
{\includegraphics[height=4cm]{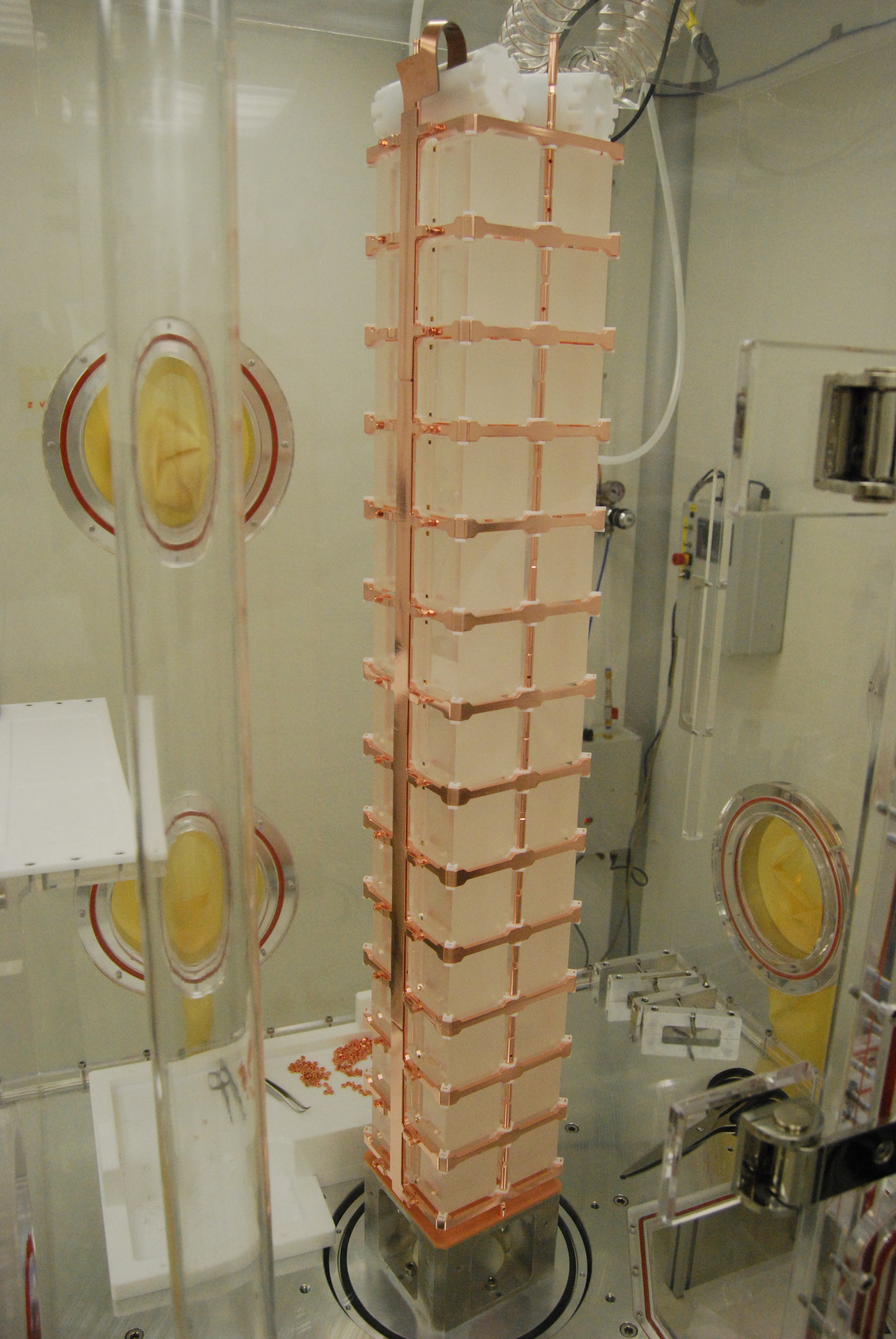}}
\hspace{.03\linewidth}
{\includegraphics[height=4cm]{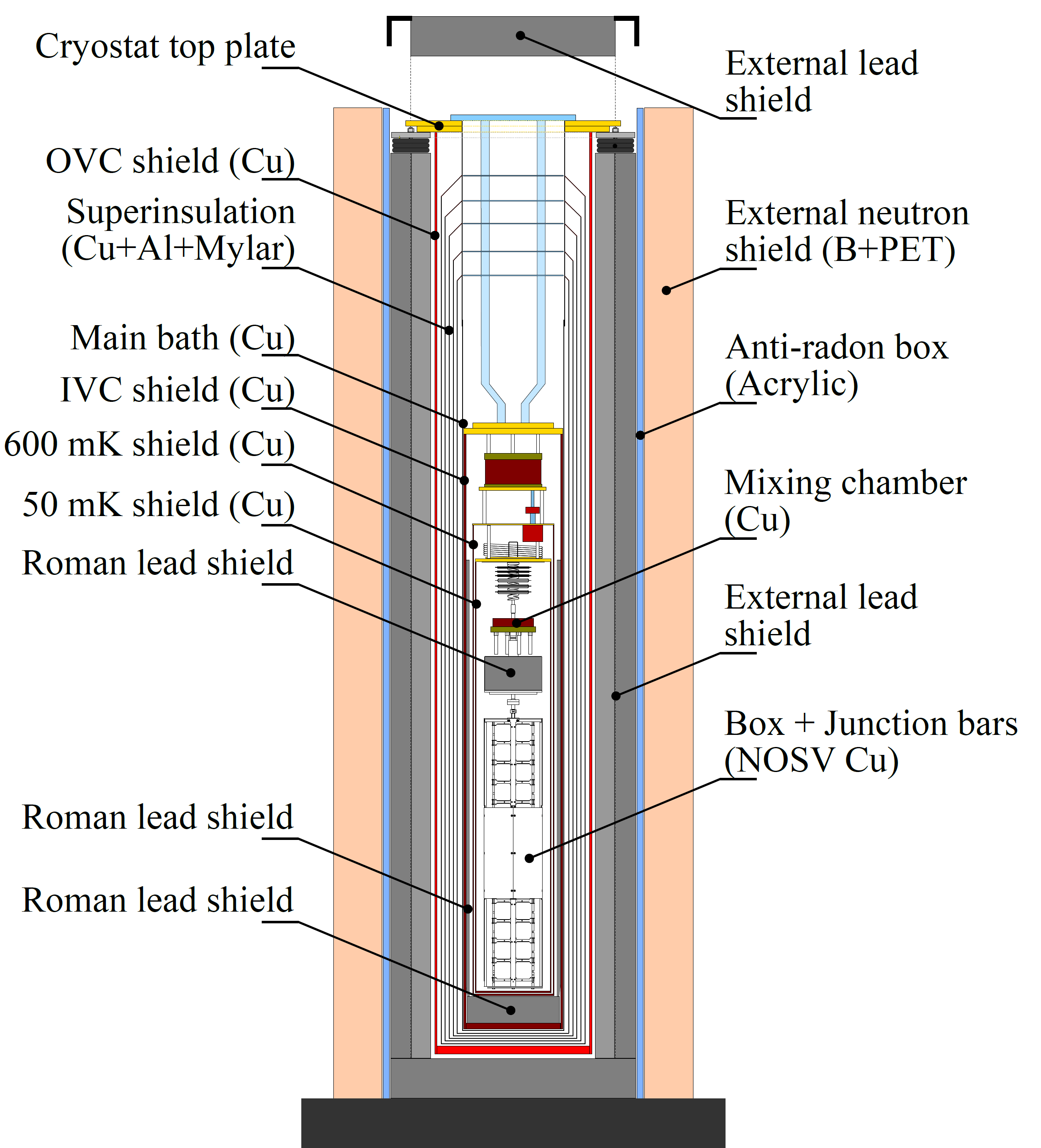}}
\hspace{.03\linewidth}
{\includegraphics[height=4cm]{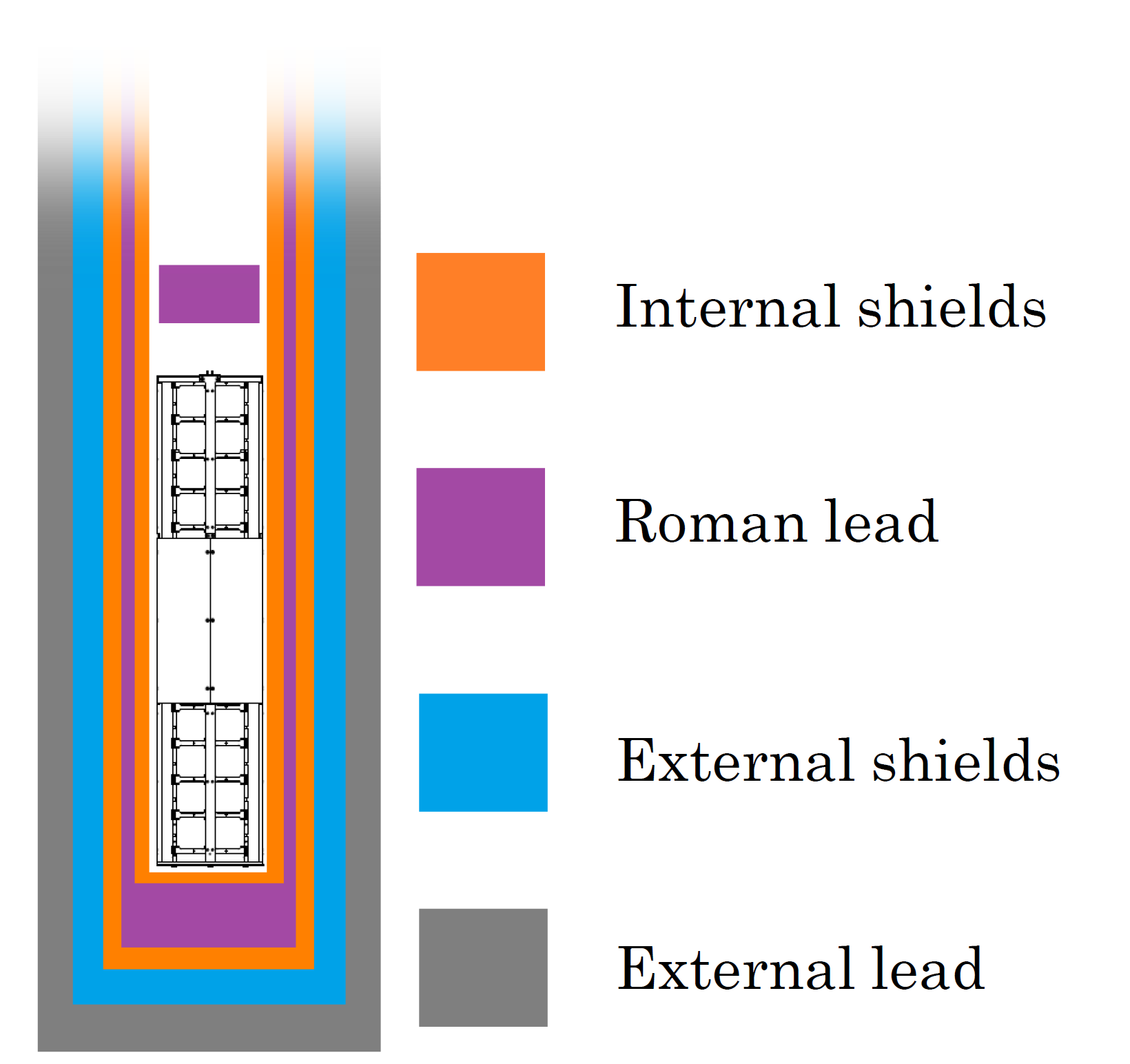}}
\caption{\footnotesize The \qz tower (left), the geometry $-$not to scale$-$ of the cryostat (center) and the scheme of the different volumes used to model the background sources in the cryostat \textit{Shields} (right).}
\label{fig:cuore0setup}
\end{center}
\end{figure}

The \qz detector is a tower of 52 \teod crystals (39~kg of total mass, 10.8~kg of \tect) operated as independent bolometers at $\sim$10~mK. At these temperatures, single particle interactions produce measurable thermal pulses proportional to the deposited energy. The modularity of the detector allows to classify the events according to their multiplicity, i.e. how many bolometers are triggered within a coincidence window of $\pm$5~ms. In particular, for the analysis of background, we build different energy spectra using the physics data collected with 33.4 kg$\cdot$y of \teod exposure:
\begin{itemize}
\item \mspecone spectrum includes the events that trigger only one bolometer (\mspecone events);
\item \mspectwo spectrum is built with the events that trigger two bolometers (\mspectwo events), using the energies deposited in each crystal;
\item \summspectwo is built with \mspectwo events, using the total energy $E$(\summspectwo) deposited in both crystals.
\end{itemize} 
The \qz tower is located in a cryostat and is surrounded by several layers of shielding to suppress the background due to environmental radioactivity (Fig.~\ref{fig:cuore0setup}). 
Based on the results of previous bolometric experiments~\cite{qino}, the main background sources are expected to be contaminations of the experimental setup. A small contribution from environmental muons and neutrons is also expected~\cite{CuoricinoMuonsCuoreExternal}. 
We simulate the background sources using a Geant4-based Monte Carlo (MC) code that generates and propagates primary and secondary particles through the \qz geometry.
We use the output of MC simulations to build the energy spectra produced in the detector by the different background sources. 
When possible, we apply some simplifications, merging the simulations of components made with the same material and, thus, characterized by identical contaminations. Similarly, we group the volumes whose contaminations produce \textit{degenerate} spectra that are indistinguishable, given the statistical uncertainty of \qz data. 
The volumes that we use as different positions for background sources are: the \teod \crystal, the \holder (i.e. the frame that supports the crystals and the surrounding cylindrical box) and four layers of \textit{Shields}.

\section{Identification of background sources}

\begin{figure}[t!]
\captionsetup[subfigure]{labelformat=empty}
\subcaptionbox{\scriptsize $\gamma$ region peaks: (1) $e^+e^-$ annihilation, (2) $^{228}$Ac, (3) $^{212}$Pb, (4) $^{212}$Bi, (5) $^{208}$Tl, (6) $^{214}$Pb, (7) $^{214}$Bi, (8) $^{40}$K, (9) $^{60}$Co.}%
[.45\linewidth]{\includegraphics[width=.45\textwidth]{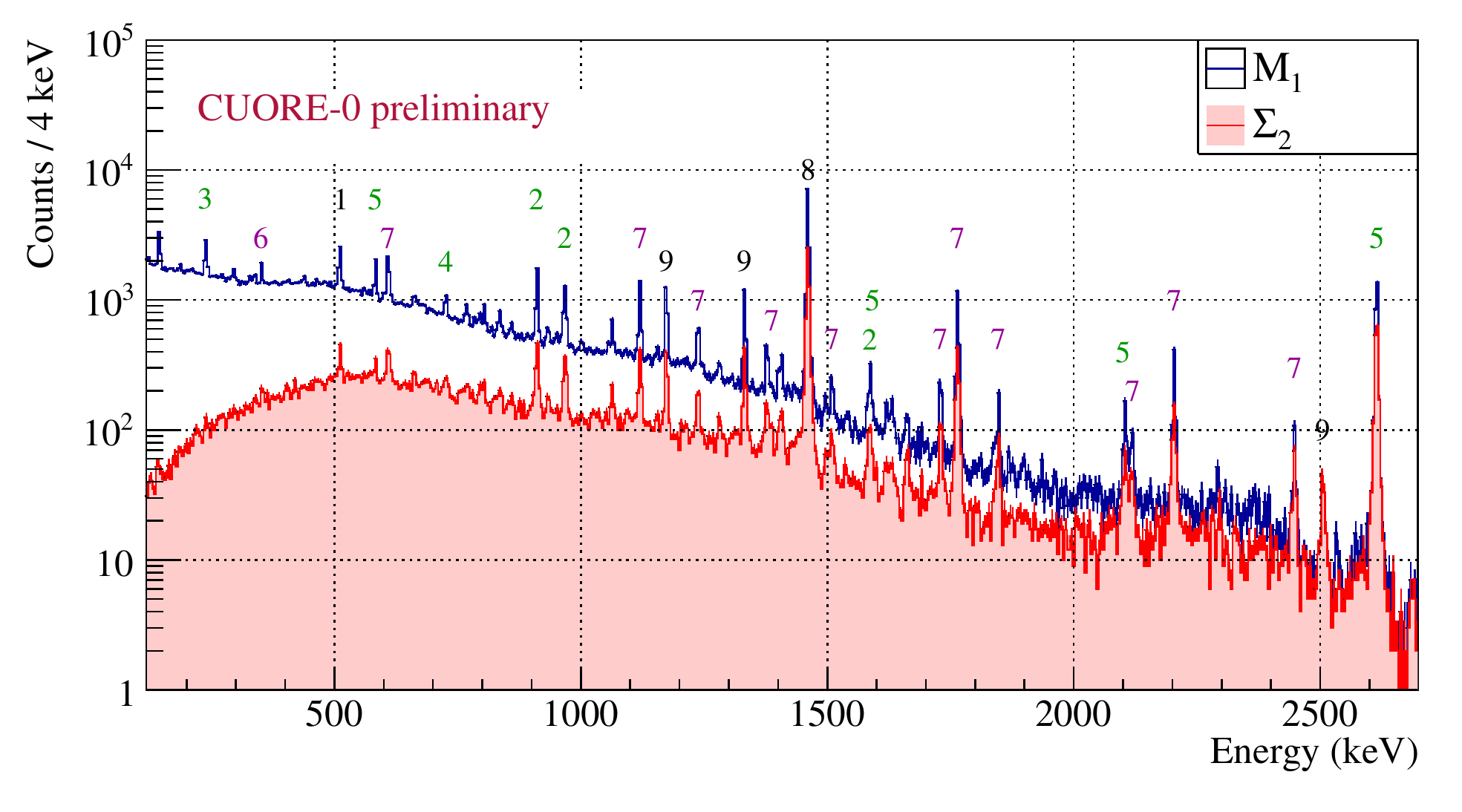}}
\hspace{.02\textwidth}
\subcaptionbox{\scriptsize $\alpha$ region peaks: (1) $^{190}$Pt, (2) $^{232}$Th, (3) $^{228}$Th, (4) $^{224}$Ra, (5) $^{220}$Rn, (6) $^{216}$Po, (7) $^{212}$Bi, (8) $^{238}$U, (9) $^{234}$U and $^{226}$Ra, (10) $^{230}$Th, (11) $^{222}$Rn, (12) $^{218}$Po, (13) $^{210}$Po.}
[.48\linewidth]{\includegraphics[width=.45\textwidth]{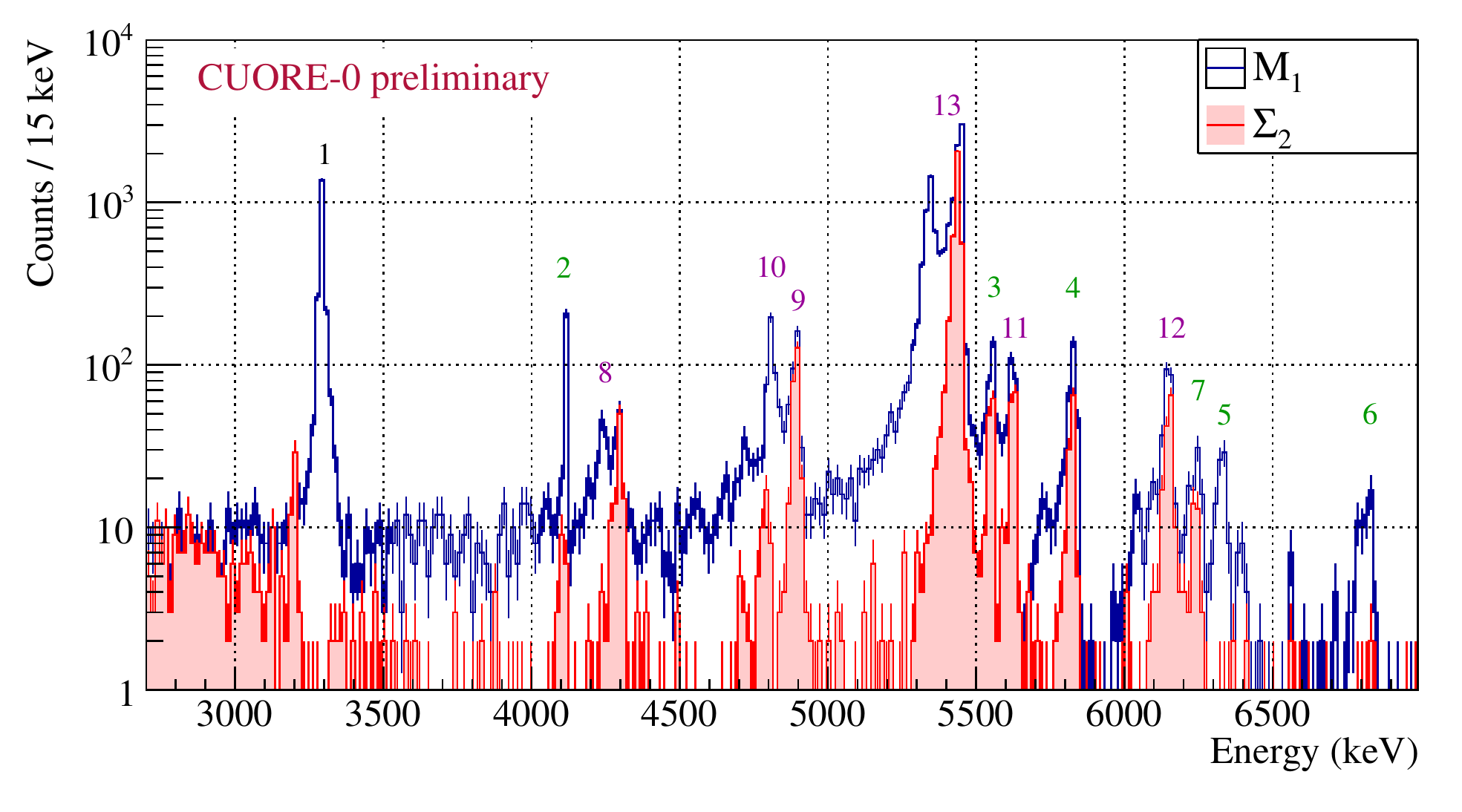}}
\caption{\footnotesize \qz \mspecone (blue) and \summspectwo (red) spectra in the $\gamma$ (left) and $\alpha$ (right) regions.}
\label{fig:PeaksID} 
\end{figure}

Identify the relevant sources to be used in the background model is crucial. Indeed, omitting a source can lead to a poor fit or to a wrong model. We exploited some \textit{a priori} information from material assays, previous bolometric experiments and cosmogenic activation calculations~\cite{Priors}. However, most of the information has been obtained from the analysis of \qz data. As shown in Fig.~\ref{fig:PeaksID}, the \mspecone and \summspectwo spectra exhibit many peaks that allow to identify the corresponding radioisotopes. Below the 2615~keV $\gamma$ line of $^{208}$Tl, the spectra include many $\gamma$ lines, while above 2615~keV most of the events are produced by $\alpha$ decays. These are the $\gamma$ and $\alpha$ regions respectively.

The peaks in the $\alpha$ region are produced by $^{190}$Pt and by radioisotopes belonging to \th and \u decay chains. Due to the short range of $\alpha$ particles and recoiling nuclei, these contaminants must be located in the \crystal or in the \holder. The analysis and the reconstruction of the $\alpha$ region is thus very useful to constrain the \th and \u activities in these components, simplifying the reconstruction of the more complicated $\gamma$ region. Thanks to the modularity of the detector, we can apply the coincidence analysis to identify the position of contaminations. For example, if an $\alpha$ decay occurs on a crystal surface, the recoil nucleus and the $\alpha$ can simultaneously interact in two adjacent crystals, producing an \mspectwo event with \summspectwo energy at the decay Q-value.
By analyzing the relative intensity of the peaks in the $\alpha$ region, we observe breaks of secular equilibrium for both \th and \u contaminations, particularly evident in the case of 5.3 and 5.4~MeV peaks of $^{210}$Po. 

In the $\gamma$ region, we observe many peaks from \th and \u decay chains and the \kq line at 1461~keV. These natural radioisotopes can be found in almost all materials, therefore the corresponding peaks can be produced by contaminations in different parts of the experimental apparatus. We also observe the peaks due to \co and other isotopes produced by cosmogenic activation or nuclear fallout (like $^{137}$Cs). In some cases, we can identify the possible positions of contaminants by comparing the relative intensities of the peaks in the experimental and simulated spectra or exploiting the available \textit{a priori} information. Otherwise, we provide to the fitting algorithm the simulated spectra of contaminations in different positions of the experimental apparatus. 

\section{Results}

The activities of the sources used for the background model are determined by fitting the experimental \mspecone, \mspectwo, and \summspectwo spectra with a linear combination of 57 simulated source spectra. 
The JAGS tool~\cite{JAGS} is used to define a Bayesian statistical model of the fit and solve it. JAGS exploits Markov Chain Monte Carlo simulations to sample the \textit{joint posterior} Probability Distribution Function (PDF) of the model parameters. The \textit{posterior} PDF is the product of the \textit{prior} and \textit{likelihood} distributions. We use the observed counts in the bins of the experimental and simulated spectra to define Poissonian \textit{likelihood}s. We define Gaussian (or half-Gaussian) \textit{prior}s when the activity of a source (or its upper limit) is known from independent measurements. Otherwise, we use uniform non-informative \textit{prior}s from 0 to upper limits higher than the maximum activities compatible with the \qz data. Finally, the \textit{joint posterior} PDF is used to evaluate the activities of the background sources and their correlations.

\begin{figure}[b!]
\centering
{\includegraphics[width=.8\textwidth]{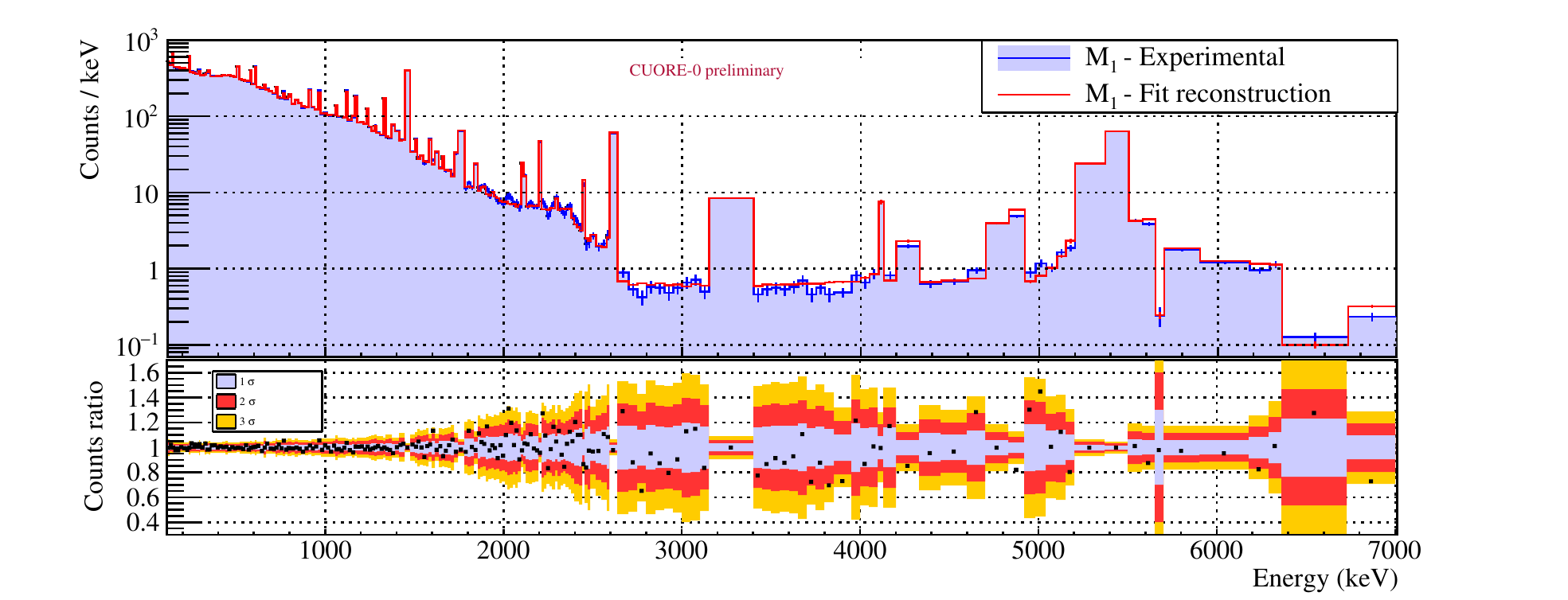}}
\caption{\footnotesize Comparison between the experimental \mspecone spectrum and JAGS reconstruction (top panel). In the bottom panel the bin by bin ratios between counts in the experimental spectrum over counts in the reconstructed one are shown; the corresponding uncertainties at 1, 2, 3 $\sigma$ are shown as colored bands centered at 1. } 
\label{Fig:JAGS-M1} 
\end{figure}

The reconstruction of the \mspecone experimental spectrum is shown in Fig.~\ref{Fig:JAGS-M1}. The normalized fit residual have a Gaussian-like distribution with mean and standard deviation compatible with 0 and 1, respectively. We obtain an equally good reconstruction of the \mspectwo and \summspectwo spectra.


\begin{figure}[t!]
\centering
{\includegraphics[width=0.45\textwidth]{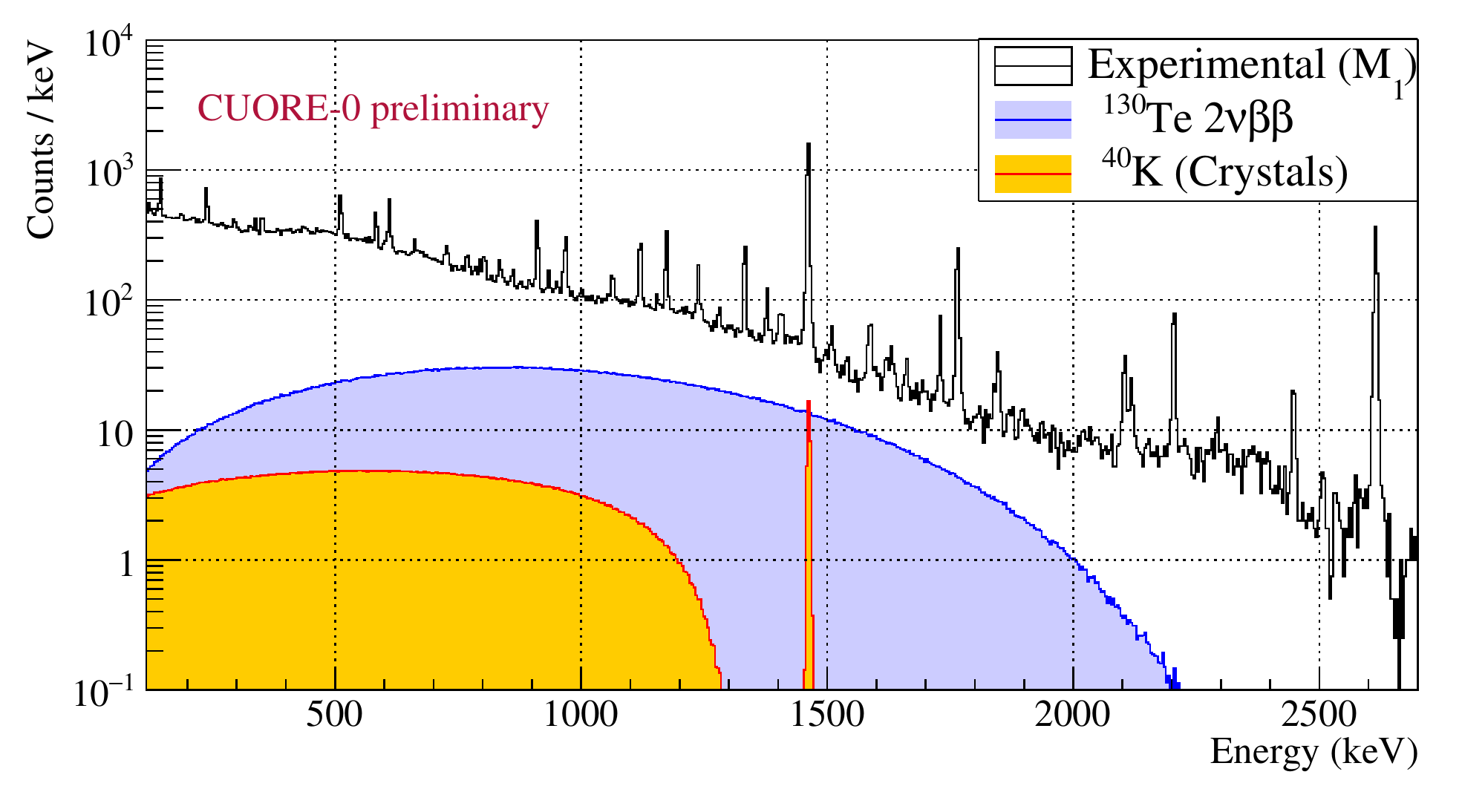}}
\hspace{.05\textwidth}
{\includegraphics[width=0.45\textwidth]{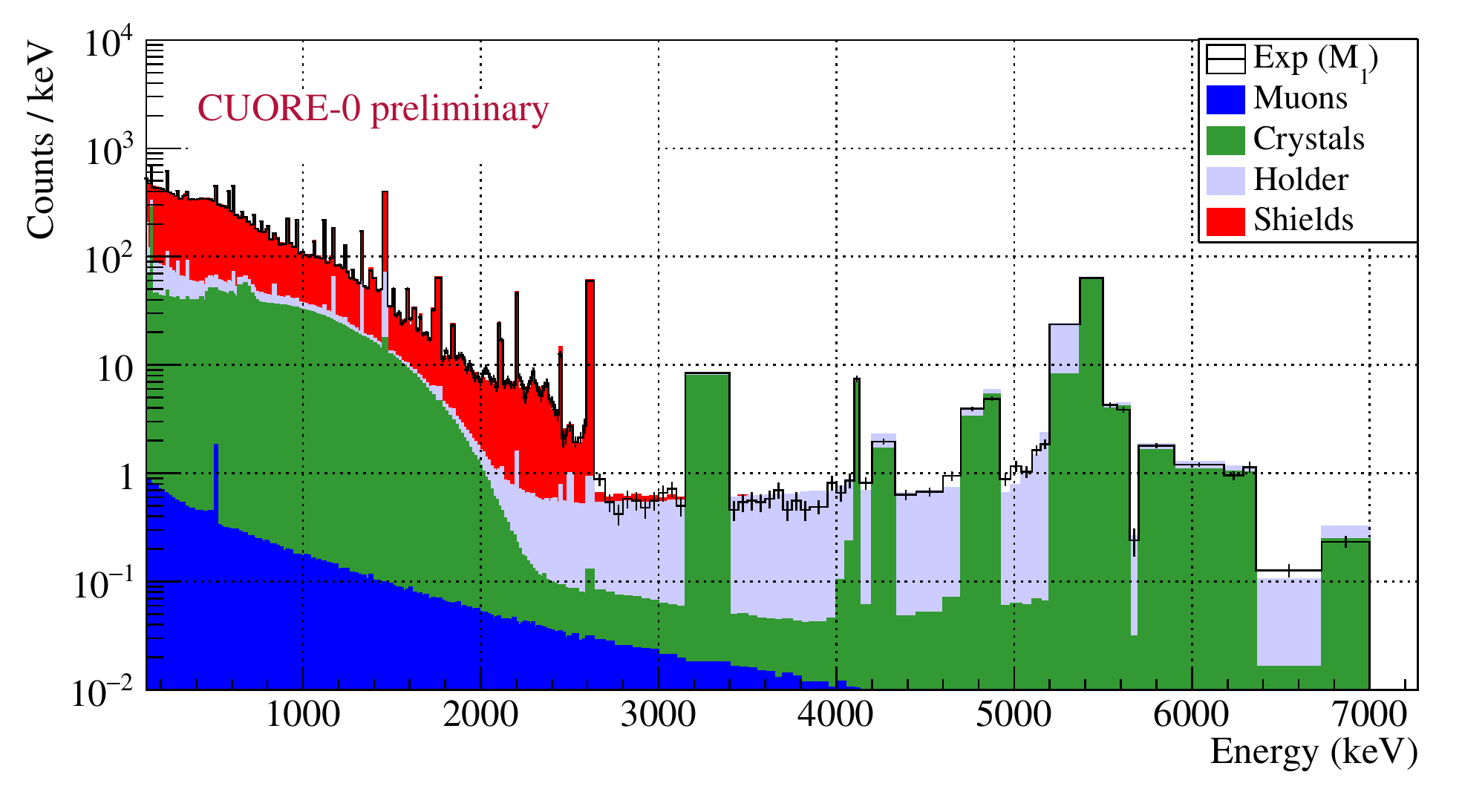}}
\caption{\footnotesize Left: CUORE-0 \mspecone spectrum compared to the evaluated contribution of \bbd and \kq background sources in \crystal.~~ Right: Background reconstruction showing the contribution of the contaminations in different positions of the experimental apparatus (stacked histogram).}	
\label{fig:2nu} 
\end{figure}

The \bbd decay of \tect produces $\sim$10\% of the events in the \mspecone $\gamma$ region from 118~keV to 2.7~MeV (Fig.~\ref{fig:2nu}, left). The resulting activity is (3.43 $\pm$ 0.09) $\times$ 10$^{−5}$~Bq/kg. The statistical uncertainty is amplified by the anti-correlation to the \kq contamination in \crystal bulk, characterized by a continuum spectrum that partially overlaps that of \bbd. 


The half-life value obtained for \bbd decay of \tect is \Td$ = [8.2 \pm 0.2 \textrm{(stat.)} \pm 0.6  \textrm{(syst.)}]\times10^{20}$y, where the systematic uncertainty was evaluated by running different fits in which the binning, energy threshold, depth of surface contaminations, priors, list of background sources, and input data were varied.


Finally, we show the reconstruction of the background produced by the contaminations in the different components of the experimental setup (Fig.~\ref{fig:2nu}, right). In the \bbz region of interest ([2470$-$2570]~keV), the largest contribution to background comes from the shields ($\sim$74\%). The \holder is the second contributor due to degraded $\alpha$s from \u and \th surface contaminants ($\sim$21\%). The remaining background is produced by \crystal contaminants and muons.

\section{Conclusions}

Through the \qz background model we evaluate the half-life of \tect \bbd decay. The preliminary result is: 
\Td$ = [8.2 \pm 0.2 \textrm{(stat.)} \pm 0.6  \textrm{(syst.)}]\times10^{20}$y.

Compared to previous results obtained from MiDBD~\cite{MIDBD2n} $[6.1 \pm 1.4 \textrm{(stat.)} ^{+2.9}_{-3.5} \textrm{(syst.)}]\times10^{20}$y and from NEMO~\cite{NEMO2n} $[7.0 \pm 0.9 \textrm{(stat.)} \pm 1.1  \textrm{(syst.)}]\times10^{20}$y, this is the most accurate measurement to date. 
We find that the background rate in the \tect \bbz region of interest is dominated by the \textit{Shields}. 
This result gives us confidence that we are on track to achieve the requirements for CUORE.

\footnotesize

\section*{Acknowledgments}

The CUORE Collaboration thanks the directors and staff of the Laboratori Nazionali del Gran Sasso and the technical staff of our laboratories. This work was supported by the Istituto Nazionale di Fisica Nucleare (INFN), the National Science, the Alfred P. Sloan Foundation, the University of Wisconsin Foundation, and Yale University. This material is also based upon work supported by the US Department of Energy (DOE) Office of Science and by the DOE Office of Science, Office of Nuclear Physics. This research used resources of the National Energy Research Scientific Computing Center (NERSC). More details can be found at: \url{http://cuore.lngs.infn.it/en/collaboration/acknowledgements}

\end{document}